\documentclass[journal]{IEEEtran}
\usepackage{xcolor}
\usepackage{soul}
\usepackage{hyperref}
\usepackage{xspace}
\usepackage{listings}
\usepackage[inline]{enumitem}
\usepackage{graphicx}
\usepackage{booktabs}
\usepackage{makecell}
\usepackage{tcolorbox}
\usepackage{balance}
\usepackage{authblk}

\newcommand{\etal}{\textit{et al.}\xspace}
\newcommand{\node}{release node\xspace}
\newcommand{\nodes}{release nodes\xspace}
\newcommand{\json}{JSON\xspace}
\newcommand{\xml}{XML\xspace}
\newcommand{\graph}{Maven Central dependency graph\xspace}
\newcommand{\nbnodes}{1,445,910\xspace}

\definecolor{codegreen}{rgb}{0,0.6,0}
\definecolor{codegray}{rgb}{0.5,0.5,0.5}
\definecolor{codelightgray}{rgb}{0.85,0.85,0.85}
\definecolor{codepurple}{rgb}{0.58,0,0.82}
\definecolor{backcolour}{rgb}{245,245,245}

\lstdefinelanguage{json}{
    basicstyle=\ttfamily\scriptsize,
    numbers=left,
    numberstyle=\tiny\color{codegray},
    stepnumber=1,
    numbersep=5pt,
    showstringspaces=false,
    breaklines=true,
    frame=single,
    backgroundcolor=\color{backcolour},
    literate=
     *{:}{{{\color{red}{:}}}}{1}
      {,}{{{\color{red}{,}}}}{1}
      {\{}{{{\color{red}{\{}}}}{1}
      {\}}{{{\color{red}{\}}}}}{1}
      {[}{{{\color{red}{[}}}}{1}
      {]}{{{\color{red}{]}}}}{1},
}

\tcbset{
    mylistingstyle/.style={
        colback=backcolour,
        colframe=black!75!black,
        boxrule=0.2pt, 
        arc=0pt,    
        boxsep=0pt 
    }
}

\lstdefinestyle{mystyle}{ 
    backgroundcolor=\color{backcolour}, 
    commentstyle=\color{codegreen},
    keywordstyle=\color{magenta},
    numberstyle=\tiny\color{codegray},
    stringstyle=\color{codepurple},
    basicstyle=\ttfamily\scriptsize,
    breakatwhitespace=false,         
    breaklines=true,                 
    captionpos=b,                    
    keepspaces=true,                 
    numbers=left,                    
    numbersep=5pt,                  
    showspaces=false,                
    showstringspaces=false,
    showtabs=false,                  
    tabsize=4,
}

\begin{document}

\providecommand*{\listingautorefname}{Listing}

\title{Software Bills of Materials in Maven Central}

\author[*]{Yogya Gamage}
\author[*]{Nadia Gonzalez Fernandez}
\author[$\dag$]{Martin Monperrus}
\author[*]{Benoit Baudry}
\affil[*]{Université de Montréal, \textit{\{yogya.gamage, nadia.gonzalez.fernandez, benoit.baudry\}@umontreal.ca}}
\affil[$\dag$]{KTH Royal Institute of Technology, \textit{monperrus@kth.se}}

\maketitle

\begin{abstract}

Software Bills of Materials (SBOMs) are essential to ensure the transparency and integrity of the software supply chain. There is a growing body of work that investigates the accuracy of SBOM generation tools and the challenges for producing complete SBOMs. Yet, there is little knowledge about how developers distribute SBOMs. In this work, we mine SBOMs from Maven Central to assess the extent to which developers publish SBOMs along with the artifacts. We develop our work on top of the Goblin framework, which consists of a \graph and a Weaver that allows augmenting the  dependency graph with additional data. For this study, we select a sample of 10\% of \nodes from the \graph and collected 14,071 SBOMs from 7,290 package releases. We then augment the \graph with the collected SBOMs. We present our methodology to mine SBOMs, as well as novel insights about SBOM publication. Our dataset is the first set of SBOMs collected from a package registry. We make it available as a standalone dataset, which can be used for future research about SBOMs and package distribution.
\end{abstract}

\IEEEpeerreviewmaketitle

\section{Introduction}

With the rise in software supply chain attacks each year, there is a growing emphasis on improving software transparency, provenance, and integrity. As a response, Software Bills of Materials (SBOMs) \cite{NTIA2021} have gained significant attention from both regulatory bodies \cite{executive_order_14028} and developers \cite{josh_2024_mcrib}. SBOMs provide a detailed list of software components and their dependencies, forming a critical foundation for applying more comprehensive security practices. 

Prior studies on SBOMs indicate a steady increase in the generation of SBOMs \cite{dalia2024sbom}, as well as in the versioning of SBOMs \cite{nocera2023software}. 
However, there is currently a lack of knowledge regarding the publication of SBOMs, along with software artifacts, in package registries. 
Yet, this is essential for application developers who fetch third-party libraries from remote registries, as it allows them to check the integrity and provenance of these libraries.

In this study, we mine SBOMs published in Maven Central and integrate them with the \graph \cite{jaime2024goblin}. To achieve this, we extend the Goblin Weaver tool, with the functionality to collect SBOM data from Maven Central. 
The \graph includes 14,459,139 \nodes, from which we sample 10\%, maintaining the original yearly distribution.
From our sample of 1,445,910 \nodes, we find 7.290 \nodes in Maven Central that are published with an SBOM, in one or multiple formats. We collect 14,071 SBOMs in total, and augment the \graph with this SBOM data.

Our analysis reveals that the 2021 White House Executive Order \cite{executive_order_14028}, which provided guidelines to improve software supply security, significantly accelerated the adoption and distribution of SBOMs, including in Maven Central. This is consistent with prior studies \cite{mirakhorli2024githubsbom}.
The CycloneDX and SPDX plugins for Gradle and Maven are the most commonly used to generate these SBOMs. 
Meanwhile, the exact process to automate the publication of SBOMs on Maven Central is not systematically documented in the SBOMs. 
We also find that most SBOMs align with the \graph. 
Yet, some SBOMs exhibit significant differences, including cases with drastically incorrect dependency information.

Our key contributions are:

\begin{itemize}

    \item An extension of the Goblin Weaver tool for collecting SBOM information from Maven Central 
    \item A new version of the \graph, augmented with SBOM data
    \item A dataset of 14,071 SBOMs collected from 7,290 releases published in Maven Central.

\end{itemize}

\section{Methodology}

In this section we introduce our process to sample nodes from the \graph and check whether the corresponding artifacts are published on Maven Central with an SBOM.

\subsection{Background}

Dependency graphs represent relationships among project components at a specific point in time. When it comes to the dependency graphs of Maven Central, nodes typically represent artifacts or releases, while edges denote dependency or version relationships. For our study, we use the Maven dependency graph dataset created by Jaime \etal \cite{jaime2024goblin} which is available publicly as a Neo4J graph database. We select the most recent dump, containing the full Maven dependency graph as of August 30, 2024, which includes a total of 15,117,217 nodes (658,078 artifact nodes and 14,459,139 \nodes). An artifact node includes the maven project coordinates (GroupID:ArtifactID) as its id. A \emph{\node} includes the release timestamp, version, and full maven coordinates, including the version (GroupID:ArtifactID:Version).

\begin{figure}
\centering
\includegraphics[width=\linewidth]{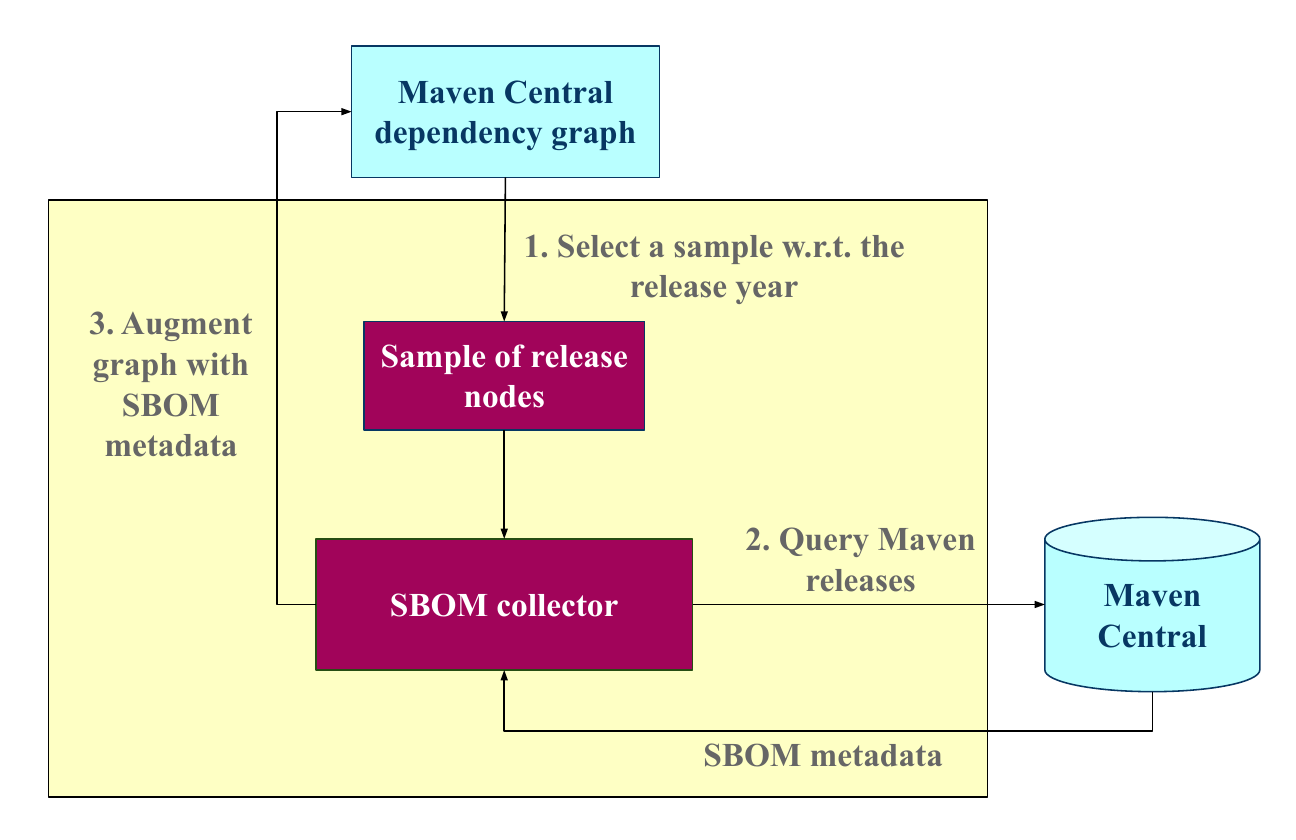}
\caption{An overview of the methodology to collect SBOM data from Maven Central and augment the \graph}
\label{fig:figure}
\end{figure}

As required by Maven Central, developers releasing a package as a JAR to Maven Central must provide the source code, Javadocs, file checksums, and signatures along with the JAR. Publishing a Software Bill of Materials (SBOM) to Maven Central is optional. However, some developers choose to include it for increasing transparency. 
In the following subsections we describe the 3 main steps of our study as illustrated in the \autoref{fig:figure}. 

\subsection{Selecting a sample of releases} \label{sec:select-sample}
For our analysis, we proportionally select representative 10\% samples of \nodes from each year in the original Neo4J dataset. This approach ensures that our sampling strategy preserves the temporal distribution of the dataset. Sampling is necessary because retrieving information for all \nodes from Maven Central is extremely time-intensive due to the volume of network requests. Based on related work \cite{nocera2023software}, our intuition is that the publication of SBOMs has evolved over time. The sample we collect and study here contains a total of \nbnodes \nodes, ranging from 1 \node in 2004 to 173,004 \nodes in 2024.


\subsection{Collecting SBOMs from Maven Central}

When searching for SBOMs in  Maven Central we focus specifically on SBOMs in \textit{CycloneDX} or \textit{SPDX}, as these are the most widely used formats \cite{balliu2023challenges}. We also search for the checksums of the SBOMs for each release. In particular, we check for the existence of an SBOM checksum in one of the following algorithms: MD5, SHA-1, SHA-256, and SHA-512.
\autoref{tab:centraldata} presents a summary of the basic information we collect from Maven Central.

In the previous step (\autoref{sec:select-sample}), we sampled a set of \nbnodes \nodes. Now, we aim at checking how many of these nodes  publish an SBOM. Using the (GroupID:ArtifactID:Version) coordinates of a node, we can construct the Maven Central URL for the \node. To determine whether a \node contains an SBOM, we retrieve its list of artifacts from the Maven Central URL. We search for files that adhere to common SBOM naming conventions, specifically those with names containing cyclonedx or spdx and file extensions \json or \xml. If such files are found, we record their URL and the SBOM format. 
Next, we search for SBOM checksums and record the checksum algorithms used.







\begin{table}[tbp]
\caption{SBOM data collected from Maven Central}
\label{tab:centraldata}
\centering

\begin{tabular}{ll}
\toprule
 Item &  Value\\
\midrule

    SBOM link &  SBOM url in Maven Central \\

    Standard & CycloneDX or SPDX \\

    Checksum algorithms & MD5,
    SHA-1, SHA-256, or SHA-512 \\

\bottomrule
\end{tabular}
\end{table}


\subsection{Augmenting the dependency graph with SBOMs}

\begin{lstlisting}[escapechar=!, language=json, caption={SBOM data as an addedValue to a release node. (The  fields added to a release node are denoted by the grey background.)}, label=lst:example-json-file, frame=none]
{
!\colorbox{codelightgray}{\parbox{0.46\textwidth}{%
\textbf{"sbom":[{"https://repo1.maven.org/maven2/org/glassfish/\\jersey/examples/https-clientserver-grizzly/2.44/\\https-clientserver-grizzly-2.44-cyclonedx.xml":"\\
        {standard=cyclonedx, 
        isSigned=true, \\
        isHashAvailable=md5, sha1}"}]}%
}}!
"id":"org.glassfish.jersey.examples:https-clientserver-grizzly:2.44",
"nodeType":"RELEASE",
"version":"2.44",
"timestamp":1722258240000}
\end{lstlisting}

To extend the Maven dependency graph with SBOM data, we use the tool GOBLIN weaver version v2.1.0 \footnote{https://github.com/Goblin-Ecosystem/goblinWeaver/tree/
65bff610455adc70ce5322a181be412465c5c9bf}. The GOBLIN weaver allows extending an existing dependency graph with new information such as SBOM in our case, without regenerating the graph from scratch. Weaver supports attaching data to a node or an edge. Since SBOMs are associated with each release, we assign the SBOM data as an added value to \nodes. 

For example, \autoref{lst:example-json-file} shows the SBOM value that we attach to the \node \textit{org.glassfish.jersey.examples:https-clientserver-grizzly:2.44}. Our implementation for augmenting the graph with SBOM data is publicly available on GitHub  \footnote{https://github.com/chains-project/sbom-mc}.

\subsection{Dataset}

\begin{figure}
\centering
\includegraphics[width=1\linewidth]{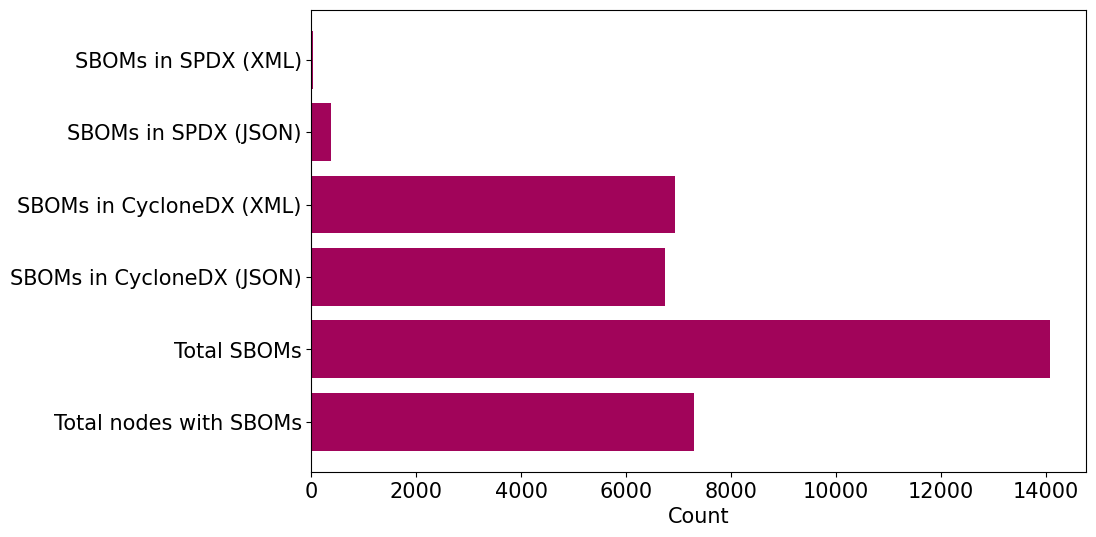}
\caption{Breakout of the 14,071 SBOMs retrieved from 7290 \nodes in Maven Central}
\label{fig:stats-dataset}
\end{figure}

 Out of 1,445,910 \nodes we queried on Maven Central, 7,290 (0.5\%) have published SBOMs, see \autoref{fig:stats-dataset}. Some releases include multiple SBOMs, often in both JSON and XML formats, and occasionally in both CycloneDX and SPDX formats. Altogether, these 7,290 \nodes over 2733 artifacts, contain a total of 14,071 SBOMs, forming our SBOM dataset that we share. 
 When considering the standards, CycloneDX is by far the most widely used, with 13,687 SBOMs; 6,748 in JSON format and 6,939 in XML format. In comparison, only 384 SBOMs are in SPDX format, with 364 in JSON and 20 in XML format. Additionally, we find that every SBOM we collect is hashed using at least one hashing algorithm. In particular, all of them are hashed with MD5, and 99\% of the SBOMs are hashed with  SHA-1 as well. Meanwhile, less than 2\% are hashed with SHA-256 or SHA-512.


We release the \graph, which we augmented with SBOM data, as a Neo4J dump \footnote{https://zenodo.org/records/10047561}. Moreover, we publish separately all the SBOM files we have collected from Maven Central.

\section{Empirical analysis of SBOMs in Maven Central}

\label{researchQuestions}

In this section, we dive into our dataset of SBOMs, looking into their evolution (RQ1), content (RQ2) and alignment with the \graph (RQ3).

\newcommand\rqCount{How many projects push an SBOM on Central?}

\newcommand\rqPatterns{How does the share of projects that publish an SBOM on Maven Central evolve over time?\xspace}

\newcommand\rqDifferences{What provenance metadata is included in SBOMs published on Maven Central?\xspace}

\newcommand\rqImpacts{Do projects that publish an SBOM have more dependents than projects that don't?\xspace}

\newcommand\rqDependencies{To what extent are  the \graph and the SBOMs aligned?\xspace}


\subsection{RQ1 \textbf{\rqPatterns}}

\begin{figure}
\centering
\includegraphics[width=0.7\linewidth]{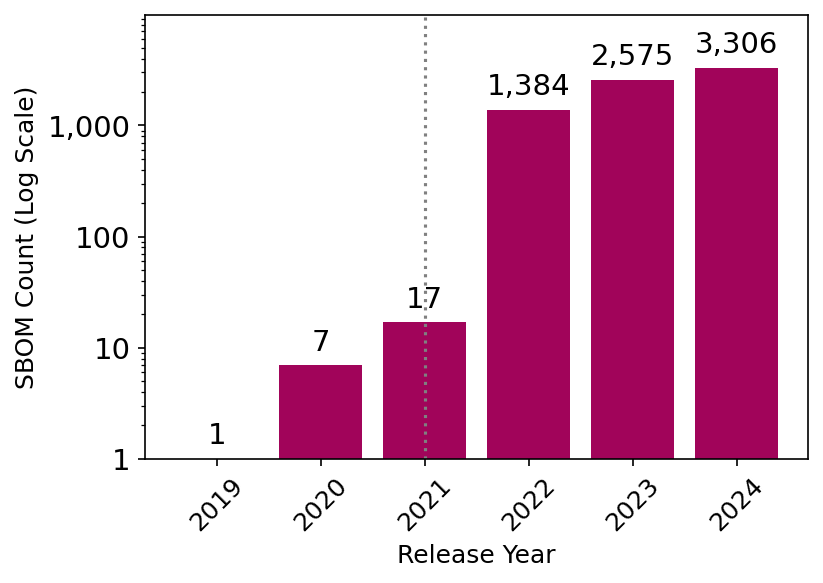}
\caption{The distribution of Maven Central \nodes that are published with an SBOM. The oldest SBOM in our dataset was released in 2019  with the \node \textit{org.spdx:spdx-tools:2.1.17}. In 2021 (dashed line), the White House published an executive order listing guidelines to improve software supply chain security, which included the publication of SBOMs.}
\label{fig:sbom-nodes}
\end{figure}

\autoref{fig:sbom-nodes} illustrates the distribution of \nodes with SBOMs over the years.
Although SBOM-generating tools were introduced as early as 2014 \cite{nocera2023software},  the first SBOM in our sample dataset is from  2019  with the \node \textit{org.spdx:spdx-tools:2.1.17} \footnote{https://repo1.maven.org/maven2/org/spdx/spdx-tools/2.1.17}. We acknowledge that the artifact org.spdx:spdx-tools first published an SBOM in 2017, but that \node is  not included in our sample. 

In 2019 and 2020, there are only 1 
and 7 
releases with SBOMs, respectively. Among the seven SBOMs published in 2020, five are associated with tools specifically designed to produce SBOMs, such as SPDX and CycloneDX. By 2021, this number increases slightly to 17, and a significant rise in SBOM publications occurrs thereafter. In 2022, 2023, and 2024, we observed 1384, 2575, and 3306 SBOMs, respectively.

This pattern suggests that while the concept of SBOMs was integrated into certain projects earlier, it took several years for developers to begin actively publishing SBOMs to Maven Central. Notably, the White House Executive Order on improving cybersecurity in 2021 likely had a significant influence, as evidenced by the sharp rise in SBOM publication after 2022, similar to observations in \cite{nocera2023software}. As of August 31, 2024, SBOM publication has increased by 1.8 times compared to 2023. By 2024, 320 different organizations such as Apache, JetBrains and GlassFish publish SBOMs to Maven Central alongside 2622 unique artifacts and 3306 total releases. 

\begin{tcolorbox}[boxrule=1pt,arc=.3em, left=4pt, right=4pt]
  \textbf{Answer to RQ1}: Despite the availability of SBOM generation tools since 2014, it took a few more years for developers to actively start publishing SBOMs to Maven Central. The number of SBOMs published on Maven Central has drastically increased after 2021, from a near absence before 2021, to thousands of published SBOMs since then. 
\end{tcolorbox}

\subsection{RQ2 \textbf{\rqDifferences}}

When analyzing the 7,290 \nodes with published SBOMs, we find that the choice of one SBOM format is usually exclusive. Only 21 \nodes include  both a CycloneDX and an SPDX SBOM. Of the total amount of \nodes, 6724 published CycloneDX SBOMs in both XML and JSON, whereas SPDX SBOMs are only published in one format. 

Typically, all CycloneDX SBOMs include SBOM creation details, such as bomFormat, version, and metadata containing author, creation tool, and timestamp information. In addition, the SBOMs include a list of components, with details such as names, publishers, hashes, licenses, and package URLs (purl). 
A component can be a library, application, operating system or a machine learning model. 
Similarly, SPDX SBOMs  include basic creation details such as SPDXID, name, and metadata. The packages section documents package creators, licenses, and summaries, while the relationships section defines the links between different SPDX documents.

For transparency, it is important that SBOMs document what tool is used to generate them. In the case of CycloneDX SBOMs the generation tool and its version are in the \textit{metadata.tools section}. In contrast, SPDX SBOMs include generation tool information under \textit{creationInfo} but omit the specific tool version. Among the 384 SPDX SBOMs, 314 are created using the \textit{spdx-gradle-plugin}, while the remaining are generated using the \textit{spdx-maven-plugin}. For CycloneDX SBOMs, 13,527 are created using the \textit{cyclonedx-maven-plugin}, and 153 with the \textit{cyclonedx-gradle-plugin}. An additional five CycloneDX SBOMs contain invalid tool names with placeholder values, and 2 CycloneDX SBOMs created in 2020, do not include a tool name. CycloneDX SBOMs generated using the Maven plugin use version 2.0.2 (released in Jul 20, 2020) or newer, with version 2.8.1 (released in Aug 04, 2024) being the most recent. Gradle plugin generated SBOMs use versions ranging between 1.7.3 (released in Dec 19, 2022) and 1.8.2 (released in Jan 19, 2024).      

By default, both the \textit{cyclonedx-maven-plugin} and \textit{spdx-maven-plugin} generate an SBOM in the target folder of the project, which can then be published to Maven Central. To publish an SBOM to Maven Central, developers can either rely on the default setup of these plugins or alternatively generate an SBOM separately and attach it as an additional artifact during the release deployment process. However, SBOMs lack systematic documentation of publication information, making it impossible to quantify the extent of usage for either publication method.  







\begin{tcolorbox}[boxrule=1pt,arc=.3em, left=4pt, right=4pt]
  \textbf{Answer to RQ2}: SPDX SBOMs are mainly generated using the Gradle version of the SPDX plugin, while the majority of the CycloneDX SBOMs are generated using the CycloneDX Maven plugin. However, SBOMs do not carry any publication information. Including publication details in SBOMs would improve their transparency and provenance.   
\end{tcolorbox}

\subsection{RQ3 \textbf{\rqDependencies}}

Here, we assess the alignment of the dependencies captured in SBOMs and the dependency relations documented in the Goblin Maven dependency graph.
We focus this analysis on the  direct dependencies of a \node, as indirect dependencies could be excluded by the developers of a project by specifying them under the exclusions section of the POM file, but this exclusion cannot be traced in the dependency graph.
Moreover, our analysis focuses on CycloneDX SBOMs, as they constitute the largest share of our dataset.

We first extract the direct dependencies listed under the dependencies tag in the CycloneDX SBOMs. Next, we query the direct dependencies of each \node from the dependency graph, comparing the results based on groupID, artifactID, and version. Since most CycloneDX SBOMs exclude test dependencies, we disregard their absence during the comparison. Aside from that, we do not evaluate the accuracy of the exact scope specified in the SBOMs.  

Our findings reveal that 11,163 out of 13,687 CycloneDX SBOMs match the dependency graph in terms of the number of direct dependencies. 
However, even among these matching SBOMs, 47 SBOMs exhibit mismatches, particularly in dependency versions. For example, while the dependency graph shows that \node \textit{org.apache.orc:orc-examples:1.7.8} depends on \textit{org.apache.hadoop:hadoop-hdfs} version 3.3.4, the corresponding SBOM lists the version as 2.2.0. 
Among the non-matching SBOMs, the largest difference occurs with the \node \textit{org.glassfish.main.extras:glassfish-embedded-web:8.0.0-M2}, where the SBOM reports 231 dependencies, while the dependency graph shows none.

One contributing factor to these mismatches is that CycloneDX SBOMs often include submodules and parent POM files as dependencies, whereas the dependency graph does not. For instance, the \node \textit{org.janusgraph:janusgraph:1.1.0-20240801-124342.ff323da} includes 20 non-test-related submodules and 4 dependencies, but the SBOM records all 24 as dependencies. 

\begin{tcolorbox}[boxrule=1pt,arc=.3em, left=4pt, right=4pt]
  \textbf{Answer to RQ3}: Most CycloneDX SBOMs align with the dependency graph when comparing direct dependencies. However, occasional misalignment occurs due to SBOMs including submodules and parent POMs as dependencies. 
\end{tcolorbox}

\section{Related work}


%
Previous work have studied the adoption of SBOMs in open-source repositories. Nocera \etal. \cite{nocera2023software} conducted a mining study on GitHub about their adoption, providing insights into the current practices and challenges in open-source software repositories. Similarly, Nordström \etal. \cite{nordstrom2024measuring} gathered  a relevant SBOM dataset to scan for vulnerabilities in the software supply chain. On the other hand, Mirakhorli \etal. \cite{mirakhorli2024githubsbom} examined both open-source and proprietary tools for SBOM generation. Most recently, \cite{zhao2024covsbom} Zhao \etal. improve the precision of SBOMs with coverage information, and Sharma \etal \cite{aman2024sbomexe} build a tool based on SBOMs to secure Java applications against dynamic class loading attacks. 
Our study complements these works, as the first work that looks into the publication of SBOMs in a package registry. 

Many tools have been developed to generate SBOMs in various formats, with previous research exploring the accuracy and usability of these tools \cite{balliu2023challenges}, \cite{halbritter2024accuracy}. Cofano \etal. and Dalia \etal. \cite{cofano2024sbom,dalia2024sbom} performed a comparative analysis of these tools and discussed that these technologies have not yet reached full automation and maturity. In RQ3, we provide novel insights about the alignment between dependencies in SBOMs and the ecosystem-scale \graph.

Scanniello \etal. and O'Donoghue \etal. \cite{scanniello2024msr4sbom,o2023impacts} investigated the differences and vulnerabilities of the SBOM generation tools SPDX and CycloneDX. They concluded that while the adoption of SBOMs is limited, there is a noticeable upward trend.
Stalnaker \etal. \cite{stalnaker2024bomsaway} conducted a comprehensive study that
investigates the current challenges stakeholders encounter when creating and using SBOMs. The paper concluded that, despite low adoption of SBOMs, practitioners utilize them for various purposes. 
Our results corroborate these previous works as we also observe a clear tendency towards more systematic publication of SBOMs along artifacts in the package registry.


\section{Conclusion}

We extend the Goblin Weaver tool to collect SBOM data from Maven Central. Through this process, we curate a dataset of 14,071 SBOMs collected from 7,290 \nodes. 
Our analysis reveals that SBOM production and documentation still face challenges, particularly in the accuracy of listed dependencies and the provenance of SBOM distribution. We envision that our dataset and the augmented dependency graph can enable deeper insights into SBOMs. For instance, these resources can be used to explore differences in SBOM content across formats and to analyze the impact of SBOMs on the popularity of a dependency.
Despite these encouraging observations about the growing adoption of SBOM publication in a package registry, we also note that SBOMs still need to spread wider in the developer's culture. In particular, we wish to stress that the recent exploration of the Subreddit r/ProgrammerHumor, yields no humor related to SBOMs \cite{kuutila2024makes}.

\balance
\bibliographystyle{IEEEtran}
\bibliography{references}

\end{document}